\documentclass[aps,pre,groupedaddress,twocolumn]{revtex4-2}

\usepackage{amsmath,amsfonts,graphicx,orcidlink}

\begin{document}

\title{The Chemotactic Index for Spatial Gradient Sensing}

\author{Nicholas A. Licata \orcidlink{0000-0001-7839-7747}}
\email[]{Contact author: licata@umich.edu}

\affiliation{Department of Natural Sciences \\ University of Michigan-Dearborn \\ Dearborn, MI 48128, USA}

\date{\today}

\begin{abstract}
We consider the problem of quantifying the chemotactic efficiency of single cells as measured by the chemotactic index $\Psi$.  Previous work in a model framework for direct sensing of spatial gradients indicated that $\Psi$ depends on a single dimensionless group $s$, which plays the role of the square of the signal to noise ratio in the problem.  We revisit this problem theoretically and demonstrate that the cumulants in the model can be calculated exactly.  We derive explicit results for the multivariate cumulants up to third order in terms of the diffusive current density and Gaunt coefficients.  We discuss the machinery required to translate Burg-Purcell style limits on concentration gradient uncertainty into results for the chemotactic index.  We compute the leading corrections to $\Psi$ in an Edgeworth expansion, and identify a dimensionless group $\lambda$ in the problem which is a ratio of concentrations that captures the effects of the non-Gaussianity.  By careful consideration of experimental results on slime mold chemotaxis, we demonstrate that the explanatory success of the original Gaussian approximation for the chemotactic index stems in part from the fact the concentration gradients were shallow, $|\lambda| \ll s$.  
\end{abstract}

\maketitle

\section{Introduction}

Chemotaxis, the directed migration of cells in response to chemical gradients, represents a fundamental biological process with profound implications across scales from bacterial navigation to immune response and embryonic development. At the heart of this phenomenon lies the remarkable ability of cells to perform spatial gradient sensing—detecting and responding to minute concentration differences across their physical dimensions, often on the order of a few percent between front and rear \cite{song2006dictyostelium}. Cells must discriminate signals from noise in environments where diffusion continuously works to dissipate the very gradients they seek to follow.  A number of different measurement strategies have evolved, which depend in part on the cellular size \cite{dusenbery1998spatial,PRXLife.2.023012}.  Small organisms like bacterial cells generally employ an indirect temporal sensing mechanism to 
measure spatial gradients in concentration \cite{mora2010limits}.  Slightly larger eukaryotic cells are able to directly sense spatial gradients of chemicals 
across their surface \cite{endres2008accuracy,hu2010physical}.  Single-celled eukaryotic cells like the slime mold {\it Dictyostelium discoideum} and the yeast {\it Sachharomyces cerevisiae} are remarkably sensitive gradient sensors \cite{fisher1989quantitative,VANHAASTERT20071787, fuller2010external}.  

Berg and Purcell were the first to 
demonstrate that diffusion sets a fundamental physical limit to the accuracy with which a cell can measure chemical concentrations \cite{berg1977physics}.  Subsequent research has highlighted the fact that the precision is also 
limited by the random binding reactions of molecules with receptors on the cell surface \cite{bialek2005physical,bialek2008cooperativity,endres2009maximum,berezhkovskii2013effect,kaizu2014berg}.  Given the remarkable chemotactic performance of cells, this begs the question whether they might be operating near fundamental physical limits \cite{mattingly2021escherichia,mattingly2026coli}.  These and related questions have spurred a cottage industry of computing modified Burg-Purcell style limits, which attempt to address these questions by layering in additional complexity which might provide a more realistic description of the problem in a living system \cite{kaizu2014berg,ten2016fundamental,aquino2016know}.  There is a growing body of literature concerning many aspects surrounding the physics of cellular concentration sensing: the role of receptors \cite{ueda2007stochastic,aquino2010increased,aquino2010increased2,aquino2011optimal,skoge2013chemical}, noise in the input signal \cite{wang2007quantifying,rappel2008receptor,govern2012fundamental,hu2014input}, and the effects of cellular 
shape \cite{PhysRevE.83.021917,tweedy2013distinct}, memory \cite{skoge2014cellular}, and communication \cite{mugler2016limits,kashyap2024trade}.  

One of the challenges related to quantitatively addressing the question of the chemotactic performance of cells is that the result of a Burg-Purcell style calculation, a concentration measurement uncertainty, is not a direct experimental observable for the system.  Instead, the implications of the cell's concentration uncertainty must be deduced and compared to what is being measured experimentally.  The chemotactic index $\Psi$ has emerged as a principal quantitative metric for characterizing the directional migration efficiency of cells \cite{ward1973chemotaxis,amselem2012control}. Defined as the ratio of the cell's net displacement toward the chemoattractant source to the total path length, $\Psi$ values near 0 indicate random migration whereas a value of 1 would indicate perfectly directed movement. This dimensionless parameter enables standardized comparison across cell types, gradient conditions, and experimental systems. However, theoretical predictions of $\Psi$ require careful translation of Burg-Purcell style limits on concentration gradient uncertainty.  This translation from concentration uncertainty to the chemotactic index is the main topic of this paper.  

We revisit this problem in the context of a model for direct gradient sensing by single cells \cite{endres2008accuracy}.  The model is simple in the sense that it treats the case of a single cell as a perfect absorber, devoid of chemical kinetics.  This simplification neglects any sources of intrinsic noise in the problem, like those associated with the production of a second-messenger \cite{ueda2007stochastic,fuller2010external}.  However, the model framework has proven to be robust enough to make accurate quantitative predictions for the chemotactic index which can be connected to experiments in certain regimes.  In the paper we will demonstrate that even in this idealized situation dominated by extrinsic noise, the behavioral response as measured by the chemotactic index will depend on more than just the signal to noise ratio.  

Previous work on direct sensing of spatial gradients demonstrated that $\Psi$ depends on a single dimensionless parameter 
\begin{equation}
s=3\pi\frac{DTR^{3}C_{z}^{2}}{C_{\infty}},
\end{equation}which plays the role of the square of the signal to noise ratio in the problem \cite{endres2008accuracy}.  From the perspective of dimensional arguments alone \cite{buckingham1914physically}, this result is somewhat surprising.  In the model framework elaborated in the next section, the cell is treated as a perfect spherical absorber.  In this case $\Psi$ [dimensionless] might depend on the cellular size $R$ [length], the averaging time $T$ [time], the diffusion coefficient of the chemoattractant $D$ [length$^{2}$ time$^{-1}$], as well as the parameters necessary to specify the external chemoattractant concentration profile (a linear ramp), hence a background concentration $C_{\infty}$ [length$^{-3}$] and a concentration gradient $C_{z}$ [length$^{-4}$].  With 6 model parameters but only 2 physical dimensions of length and time, we might expect $6-2=4$ dimensionless groups in the problem.  However, $\Psi$ and $s$ are only two dimensionless groups, so what are the others?  In this paper we partially resolve the riddle of the missing dimensionless groups, by demonstrating that the chemotactic index also depends on another dimensionless group which is a ratio of concentrations,
\begin{equation}
\lambda = \frac{RC_{z}}{C_{\infty}}. \label{lambda}
\end{equation}We demonstrate that the statistical quantities characterizing the model can be determined exactly by computing the cumulant generating function.  From these cumulants, we approximate the marginal distribution necessary to compute the chemotactic index.  By calculating the higher order cumulants in the model, we demonstrate that the original result $\Psi=\Psi(s)$ (Eq. \ref{Psigauss}) depends on a Gaussian approximation which is only valid in the limit of shallow gradients, $| \lambda | \ll s$. We use an explicit expression for the third order cumulants to compute the leading corrections to $\Psi=\Psi(s,\lambda)$ in terms of an Edgeworth expansion.  We conclude by discussing the relationship of the results to the experimental observations.

\section{Spatial Gradient Sensing}

In this section we review the theory of spatial gradient sensing of Endres and Wingreen \cite{endres2008accuracy}.  The essential results of this section are the same as in the original paper, although the notation is slightly different, and the method is presented in a way that emphasizes the importance of the spherical harmonics, which will become useful in the next section.  Consider a spherical cell of radius $R$ in the presence of a chemical with diffusivity $D$.  Let $(r,\theta,\phi)$ denote a spherical polar coordinate system whose origin coincides with the center of the cell. The idealized {\it perfectly absorbing sphere} has access to the locations ${\bf r}_{j}=(R,\theta_j,\phi_j)$ and absorption times $t_{j}$ for the $N$ particles absorbed during the time interval $T$.  For the cell, the goal is to best estimate the gradient of the chemical species from the absorption events.  

An estimate of both the background concentration $C_{\infty}$ and the background ($B$) concentration gradient $\boldsymbol{\nabla} C_{B} = C_{x} {\bf \hat{x}}+C_{y} {\bf \hat{y}}+C_{z} {\bf \hat{z}}$ can be obtained by performing a linear regression 
of the instantaneous current density to the average current density impinging on the spherical surface.  Technically, this amounts to minimizing the resulting error $E$, $\frac{\partial E}{\partial C_{k}}=0$, with respect to the background 
concentration parameters $C_{k}$ for each $k \in \{\infty,x,y,z\}$, 
\begin{equation}
E=\int_{0}^{T}\mathrm{d}t \iint \mathrm{d}A \left( \sum_{j=1}^{N} \delta(t-t_{j})\delta^{2}(\mathbf{r}-\mathbf{r}_{j})-J(\theta,\phi,t)\right)^2.
\end{equation}
Here the area element $\mathrm{d}A=R^{2}\mathrm{d}\Omega = R^{2}\sin\theta\mathrm{d}\theta\mathrm{d}\phi$.  The average current density impinging on the spherical surface 
$J(\theta,\phi,t)=-D\boldsymbol{\nabla} C(\mathbf{r},t) \cdot \left. (-\mathbf{\hat{r}})\right|_{r=R}$ is calculated from the solution for the concentration field $C(\mathbf{r},t)$.  
Introducing dimensionless variables for length $\boldsymbol{\xi}=\mathbf{r}/R$, time $\tau=\frac{Dt}{R^{2}}$, and concentration $c=CR^{3}$, the governing equation for the concentration is the diffusion equation, 
\begin{equation}
\frac{\partial c}{\partial \tau} = \nabla^{2}c.
\end{equation}
Here it is understood that the radial derivatives in the Laplacian are with respect to $\xi$.  
The theoretical description assumes the case of steady-state diffusion, $\frac{\partial c}{\partial \tau}=0$, in which case the concentration is a solution to Laplace's equation $\nabla^{2}c=0$.  The solution which satisfies the boundary conditions of vanishing concentration at 
the surface of the cell, $c(\xi=1,\theta,\phi)=0$, and has the correct far-field behavior is
\begin{equation}
c(\xi,\theta,\phi) = c_{\infty}\left(1-\frac{1}{\xi}\right) + \left( \xi - \frac{1}{\xi^{2}}\right) \boldsymbol{\nabla}c_{B} \cdot \boldsymbol{\hat{\xi}}.
\end{equation}
Defining the dimensionless background concentration parameters $c_{\infty}=R^{3}C_{\infty}$, $c_{x}=R^{4}C_{x}$, $c_{y}=R^{4}C_{y}$, $c_{z}=R^{4}C_{z}$, and $c_{\pm}=(c_{x}\pm i c_{y})/2$ we can write  
\begin{equation}
 \boldsymbol{\nabla}c_{B} \cdot \boldsymbol{\hat{\xi}} =  c_{-}\sin\theta e^{i\phi} + c_{+}\sin\theta e^{-i\phi} + c_{z}\cos\theta.  
\end{equation}
The dimensionless error $\mathcal{E}=\frac{R^4E}{D}$ is then 
\begin{equation}
\mathcal{E}=\int_{0}^{\Delta} \mathrm{d}\tau \iint \mathrm{d}\Omega \left( \sum_{j=1}^{N} \delta(\tau - \tau_{j})\delta^{2}(\boldsymbol{\xi}-\boldsymbol{\xi}_{j}) - j(\theta,\phi,\tau) \right)^{2}, \label{dimE}
\end{equation}
where $\Delta=\frac{DT}{R^2}$.  To perform the error minimization, it is useful to expand the current density and the angular delta function in terms of spherical harmonics $Y_{\ell m}(\theta,\phi)$, 
\begin{equation}
j(\theta,\phi,\tau) = \left.\frac{\partial c}{\partial \xi}\right|_{\xi=1} = \sum_{\ell=0}^{\infty}\sum_{m=-\ell}^{\ell}j_{\ell m}(\tau)Y_{\ell m}(\theta,\phi)
\end{equation}
and
\begin{equation}
\delta^{2}(\boldsymbol{\xi}-\boldsymbol{\xi}_{j}) = \frac{1}{\xi_{j}^{2}} \sum_{\ell=0}^{\infty}\sum_{m=-\ell}^{\ell} Y^{*}_{\ell m}(\theta_{j},\phi_{j}) Y_{\ell m}(\theta,\phi).  
\end{equation} 
Note that $\xi_{j}=1$ since all particle arrivals are at the surface of the sphere.  In a more general setting the coefficients that appear in the expansion for the current density might depend on time, but in this steady-state model they are constants, $j_{\ell m}(\tau)=j_{\ell m}$.  In fact, there are only four non-zero $j_{\ell m}$ in the model, $j_{00}=c_{\infty}\sqrt{4\pi}$, $j_{10}=c_{z}\sqrt{12\pi}$, $j_{11}=-c_{-}\sqrt{24\pi}$, and $j_{1,-1}=c_{+}\sqrt{24\pi}$.  The error minimization is performed by solving $\frac{\partial\mathcal{E}}{\partial \alpha}=0$ for each $\alpha \in \{c_{\infty},c_{+},c_{-},c_{z}\}$,  
\begin{widetext}
\begin{equation}
\int_{0}^{\Delta} \mathrm{d}\tau \iint \mathrm{d}\Omega \sum_{\ell=0}^{\infty} \sum_{m=-\ell}^{\ell} \left( \sum_{j=1}^{N} \delta(\tau-\tau_{j}) Y_{\ell m}^{*}(\theta_{j},\phi_{j}) - j_{\ell m}\right) Y_{\ell m}(\theta,\phi)\sum_{L=0}^{\infty} \sum_{M=-L}^{L}\frac{\partial j_{LM}}{\partial \alpha}Y_{LM}(\theta,\phi) = 0.
\end{equation}
\end{widetext}
Using the relation $Y_{LM}(\theta,\phi)=(-1)^{-M}Y_{L,-M}^{*}(\theta,\phi)$ and the orthonormality of the spherical harmonics, $\iint Y_{LM}^{*}(\theta,\phi)Y_{\ell m}(\theta, \phi)\mathrm{d}\Omega=\delta_{L,\ell}\delta_{M,m}$, we can perform the angular integration to find
\begin{widetext}
\begin{equation}
\int_{0}^{\Delta} \mathrm{d}\tau \sum_{\ell m} \left( \sum_{j=1}^{N} \delta(\tau-\tau_{j}) Y_{\ell m}^{*}(\theta_{j},\phi_{j}) - j_{\ell m}\right) (-1)^{m}\frac{\partial j_{\ell,-m}}{\partial \alpha} = 0.
\end{equation}
\end{widetext}
Here we have introduced the shorthand notation $\sum_{\ell m} = \sum_{\ell=0}^{\infty}\sum_{m=-\ell}^{\ell}$.  
The Kronecker delta $\delta_{L,\ell}=1$ if $L=\ell$ and $\delta_{L,\ell}=0$ if $L \neq \ell$.  
For the derivative with respect to the background concentration we compute $\frac{\partial j_{\ell m}}{\partial c_{\infty}}=\sqrt{4\pi}\delta_{\ell,0}\delta_{m,0}$.  The temporal integration is easily performed using the delta function, since all of the particle arrivals occurred during the time interval $0 \leq \tau_{j} \leq \Delta$.  We also use the fact that the $j_{\ell m}$ are independent of time.  This results in an equation for the best estimate of the background concentration, denoted by a tilde over the variable,  
\begin{equation}
\tilde{c}_{\infty} = \frac{1}{\Delta\sqrt{4\pi}}\sum_{j=1}^{N}Y_{00}^{*}(\theta_{j},\phi_{j}).  
\end{equation}
For the derivatives with respect to the gradient component $c_{z}$ we have $\frac{\partial j_{\ell m}}{\partial c_{z}}=\sqrt{12\pi}\delta_{\ell,1}\delta_{m,0}$ and hence \\
\begin{equation}
\tilde{c}_{z} = \frac{1}{\Delta\sqrt{12\pi}}\sum_{j=1}^{N}Y_{10}^{*}(\theta_{j},\phi_{j}).  
\end{equation}
For $c_{+}$ we have $\frac{\partial j_{\ell m}}{\partial c_{+}}=\sqrt{24\pi}\delta_{\ell,1}\delta_{m,-1}$ and hence \\
\begin{equation}
\tilde{c}_{-} = \frac{-1}{\Delta\sqrt{24\pi}}\sum_{j=1}^{N}Y_{11}^{*}(\theta_{j},\phi_{j}).  
\end{equation}
For $c_{-}$ we have $\frac{\partial j_{\ell m}}{\partial c_{-}}=-\sqrt{24\pi}\delta_{\ell,1}\delta_{m,1}$ and hence 
\begin{equation}
\tilde{c}_{+} = \frac{1}{\Delta\sqrt{24\pi}}\sum_{j=1}^{N}Y_{1,-1}^{*}(\theta_{j},\phi_{j}).  
\end{equation}
The relations for $\tilde{c}_{+}$ and $\tilde{c}_{-}$ can be combined to obtain the best estimates for the Cartesian components of the gradient in the $x$ and $y$ directions.
\begin{eqnarray}
\tilde{c}_{x} =   \frac{1}{\Delta\sqrt{24\pi}}\sum_{j=1}^{N}\left( Y_{1,-1}^{*}(\theta_{j},\phi_{j}) - Y_{11}^{*}(\theta_{j},\phi_{j}) \right) \\
\tilde{c}_{y} =   \frac{-i}{\Delta\sqrt{24\pi}}\sum_{j=1}^{N}\left( Y_{1,-1}^{*}(\theta_{j},\phi_{j}) + Y_{11}^{*}(\theta_{j},\phi_{j}) \right) 
\end{eqnarray} 
Ultimately we are interested in the sensitivity of the gradient sensing, as quantified by the variance of the best estimates.  The best estimates depend on the number $N$ and locations $\{ \theta_{j},\phi_{j}\}_{j=1,2,...,N}$ of the particle arrivals during the time interval $\Delta$.  As a result we must average over the arrival distribution.  For this Poisson process, averages can be computed in the following fashion (see Appendix A in \cite{bialek2012biophysics}),
\begin{widetext}
\begin{equation}
\langle \mathcal{F}\rangle = e^{-\int_{0}^{\Delta}r(\tau)\mathrm{d}\tau}\sum_{N=0}^{\infty}\frac{1}{N!}\int_{0}^{\Delta}\mathrm{d}\tau_{1}\cdots\int_{0}^{\Delta}\mathrm{d}\tau_{N}\left( \prod_{j=1}^{N} r(\tau_{j}) \right) \mathcal{F}(\{\theta_{i},\phi_{i},\tau_{i}\}_{i=1,2,...,N}).
\end{equation}
\end{widetext}
This expression can be simplified, because in the steady-state model, the Poisson process is homogeneous, and the rate parameter is constant, 
\begin{equation}
r(\tau)=\iint \mathrm{d}\Omega \, j(\theta,\phi,\tau) = \sqrt{4\pi} j_{00} = 4\pi c_{\infty}.
\end{equation}
The average number of particle arrivals in the time interval is then simply $\langle N \rangle=\int_{0}^{\Delta} r(\tau) \mathrm{d}\tau  = 4\pi c_{\infty} \Delta$.  Note that from the equations for the best estimates of the concentration gradient, the quantities to be averaged are independent of time as well, so $\mathcal{F}(\{\theta_{i},\phi_{i},\tau_{i}\}_{i=1,2,...,N})=\mathcal{F}(\{\theta_{i},\phi_{i}\}_{i=1,2,...,N})$.  

At this point we can state one of the major results of \cite{endres2008accuracy}, which is a quantification of the gradient sensing uncertainty from the computed variances of the best estimators.  Denoting the variance of $\mathcal{F}$ by $\text{var}(\mathcal{F})=\langle \mathcal{F}^{2} \rangle-\langle \mathcal{F} \rangle^{2}$, Endres and Wingreen found that 
\begin{equation}
\text{var}(\tilde{c}_{\infty}) = \frac{c_{\infty}}{4\pi \Delta}
\end{equation}
and 
\begin{equation}
\text{var}(\tilde{c}_{x})=\text{var}(\tilde{c}_{y})=\text{var}(\tilde{c}_{z})=\frac{c_{\infty}}{12\pi\Delta}.
\end{equation}
Somewhat remarkably, the results are independent of the actual gradient $(c_{x},c_{y},c_{z})$ entirely, as the only concentration dependence is on the background concentration $c_{\infty}$.  
\section{Cumulant Generating Function}

Before moving on to discuss how the theoretical results might be connected to experiments through the chemotactic index, it is worth pausing to consider the statistical structure of the model in more detail, as these questions are 
closely related.  In the original work, apart from mentioning that the components of the gradient are independent, there is no discussion regarding the cross-moments, for example, 
\begin{equation}
\langle \tilde{c}_{\infty} \tilde{c}_{z}\rangle - \langle \tilde{c}_{\infty} \rangle \langle \tilde{c}_{z} \rangle = \frac{c_{z}}{4\pi\Delta},
\end{equation}
or higher order cumulants, for example, 
\begin{equation}
\langle \tilde{c}_{z}^{3} \rangle - 3 \langle \tilde{c}_{z}^{2} \rangle \langle \tilde{c}_{z} \rangle + 2 \langle \tilde{c}_{z} \rangle^{3} = \frac{3c_{z}}{80\pi^{2}\Delta^{2}}.
\end{equation}
As we will see in the next section, the implicit assumptions of the original calculation in \cite{endres2008accuracy} of the chemotactic index was that the higher order cumulants were all zero, which they are not.  By calculating them, we can test the regime of validity for the approximation in which they are neglected.  

To begin we define a concentration vector $\boldsymbol{\mathcal{C}}= (\mathcal{C}^{\infty},\mathcal{C}^{+},\mathcal{C}^{-},\mathcal{C}^{z})$ where the components $\mathcal{C}^{i}$ (super-scripts denote components) are defined by the following rescalings of the best estimates,
\begin{eqnarray}
\mathcal{C}^{\infty} &=& \tilde{c}_{\infty}\Delta \sqrt{4\pi}  = \sum_{i=1}^{N}Y_{00}^{*}(\theta_{i},\phi_{i}) \label{Cinftybest} \\ 
\mathcal{C}^{+} &=& \tilde{c}_{+}\Delta \sqrt{24\pi}  = \sum_{i=1}^{N}Y_{1,-1}^{*}(\theta_{i},\phi_{i}) \label{Cplusbest} \\ 
\mathcal{C}^{-} &=& -\tilde{c}_{-} \Delta \sqrt{24\pi}  = \sum_{i=1}^{N}Y_{11}^{*}(\theta_{i},\phi_{i}) \label{Cminusbest}\\ 
\mathcal{C}^{z} &=& \tilde{c}_{z}\Delta \sqrt{12\pi}  = \sum_{i=1}^{N}Y_{10}^{*}(\theta_{i},\phi_{i})  \label{Czbest}
\end{eqnarray}
Defining the vector $\boldsymbol{\zeta}=(\zeta^{\infty},\zeta^{+},\zeta^{-},\zeta^{z})$ the multivariate moment generating function can be written as 
\begin{equation}
M_{\boldsymbol{\mathcal{C}}}(\boldsymbol{\zeta}) = \langle e^{\zeta_{i}\mathcal{C}^{i}}\rangle
\end{equation}
where it is understood that repeated indices are summed over using the Einstein summation convention, with the index $i \in \{\infty,+,-,z\}$.  Since each index $i$ is associated with a single spherical harmonic indexed by $L$ and $M$, notationally it is useful to make the identification ($i=LM$), where ($\infty=00$), ($+=1,-1$), ($-=11$), and ($z=10$).  Then we can write 
\begin{equation}
M_{\boldsymbol{\mathcal{C}}}(\boldsymbol{\zeta}) = \langle e^{\sum_{\ell m}\zeta_{\ell m}\sum_{j=1}^{N}Y_{\ell m}^{*}(\theta_{j},\phi_{j})}\rangle.
\end{equation}
Performing the averaging we find 
\begin{widetext}
\begin{eqnarray}
M_{\boldsymbol{\mathcal{C}}}(\boldsymbol{\zeta}) &=& e^{-\langle N \rangle} \sum_{N=0}^{\infty}\frac{1}{N!}\int_{0}^{\Delta}\mathrm{d}\tau_{1}r(\tau_{1})\cdots \int_{0}^{\Delta}\mathrm{d}\tau_{N}r(\tau_{N}) e^{\sum_{j=1}^{N}\sum_{\ell m}\zeta_{\ell m}Y_{\ell m}^{*}(\theta_{j},\phi_{j})} \\
&=& e^{-\langle N \rangle}  \sum_{N=0}^{\infty}\frac{1}{N!}\int_{0}^{\Delta}\mathrm{d}\tau_{1}r(\tau_{1})\cdots \int_{0}^{\Delta}\mathrm{d}\tau_{N}r(\tau_{N}) \prod_{j=1}^{N} e^{\sum_{\ell m}\zeta_{\ell m}Y_{\ell m}^{*}(\theta_{j},\phi_{j})} \\
&=& e^{-\langle N \rangle}  \sum_{N=0}^{\infty}\frac{1}{N!} \left( \int_{0}^{\Delta} \mathrm{d}\tau_{1}r(\tau_{1})e^{\sum_{\ell m}\zeta_{\ell m}Y_{\ell m}^{*}(\theta_{1},\phi_{1})}\right)^{N} \\
&=& e^{-\langle N \rangle} e^{\int_{0}^{\Delta}\mathrm{d}\tau\, r(\tau)e^{\sum_{\ell m}\zeta_{\ell m}Y_{\ell m}^{*}(\theta,\phi)}} \\
&=& e^{\int_{0}^{\Delta}\mathrm{d}\tau \,r(\tau) \left( e^{\sum_{\ell m}\zeta_{\ell m}Y_{\ell m}^{*}(\theta,\phi)}-1\right)}.  
\end{eqnarray}
The cumulant generating function is then computed as 
\begin{eqnarray}
K_{\boldsymbol{\mathcal{C}}}(\boldsymbol{\zeta}) = \log (M_{\boldsymbol{\mathcal{C}}}(\boldsymbol{\zeta}))=\Delta \sum_{\ell' m'}j_{\ell' m'}\iint \mathrm{d}\Omega \, Y_{\ell'm'}(\theta,\phi) \left( \sum_{\ell m}\zeta_{\ell m}Y_{\ell m}^{*}(\theta,\phi) + \frac{1}{2!} \sum_{\ell m}\zeta_{\ell m}Y_{\ell m}^{*}(\theta,\phi) \sum_{r s}\zeta_{r s}Y_{r s}^{*}(\theta,\phi)\right.   \nonumber \\
\left. + \frac{1}{3!} \sum_{\ell m}\zeta_{\ell m}Y_{\ell m}^{*}(\theta,\phi) \sum_{r s}\zeta_{r s}Y_{r s}^{*}(\theta,\phi) \sum_{pq}\zeta_{pq}Y_{pq}^{*}(\theta,\phi) + \cdots \right).
\label{cumulantgenerating}
\end{eqnarray}
\end{widetext}
The cumulants can be obtained by differentiating $K_{\boldsymbol{\mathcal{C}}}(\boldsymbol{\zeta}) $, since 
\begin{equation}
K_{\boldsymbol{\mathcal{C}}}(\boldsymbol{\zeta}) =\zeta_{i}\kappa^{i} + \frac{1}{2!}\zeta_{i}\zeta_{j}\kappa^{i,j} + \frac{1}{3!}\zeta_{i}\zeta_{j}\zeta_{k}\kappa^{i,j,k} + \cdots \label{cumulantgenerating2}
\end{equation}
The multivariate non-central moment 
\begin{equation}
\kappa^{j}=\langle \mathcal{C}^{j} \rangle = \left.\frac{\partial K_{\boldsymbol{\mathcal{C}}}(\boldsymbol{\zeta})}{\partial \zeta_{j}}\right|_{\boldsymbol{\zeta}=\boldsymbol{0}}.
\end{equation}
The resulting integral can be computed directly from Eq. (\ref{cumulantgenerating}) using the orthonormality of the spherical harmonics.  Using the fact that $\zeta_{i}=\zeta_{LM}$ and $\frac{\partial \zeta_{\ell m}}{\partial \zeta_{LM}} = \delta_{\ell,L}\delta_{m,M}$ we find
\begin{equation}
\kappa^{i}= \Delta \sum_{\ell'm'}j_{\ell'm'}\sum_{\ell m}\delta_{\ell,\ell'}\delta_{m,m'}\frac{\partial \zeta_{\ell m}}{\partial \zeta_{LM}}=\Delta j_{LM}.  
\end{equation}
For example, if $i=z$ then $LM=10$ and $\kappa^{z}=c_{z} \Delta \sqrt{12\pi}$.  From Eq. (\ref{Czbest}) we conclude $\langle \tilde{c}_{z}\rangle = c_{z}$ as expected.  

The multi-variate non-central moment
\begin{equation}
\kappa^{ij}=\langle \mathcal{C}^{i}\mathcal{C}^{j} \rangle
\end{equation}
is related to the covariance matrix as 
\begin{equation}
\kappa^{i,j} = \kappa^{ij}-\kappa^{i}\kappa^{j}.
\end{equation}
The use of a comma separating the indices indicates a cumulant, instead of a moment.  The inverse of the covariance matrix is indicated by downstairs indices, $(\kappa^{-1})^{i,j}=\kappa_{i,j}$.  The covariance matrix is computed as 
\begin{equation}
\kappa^{i,j} = \left.\frac{\partial^{2} K_{\boldsymbol{\mathcal{C}}}(\boldsymbol{\zeta})}{\partial \zeta_{i} \partial \zeta_{j}}\right|_{\boldsymbol{\zeta}=\boldsymbol{0}}.
\end{equation}
The computation requires the integral of a product of three spherical harmonics (note two of them are actually the complex conjugates), which can be expressed in terms of the Gaunt coefficient \cite{sebilleau1998computation}
\begin{widetext}
\begin{equation}
\mathsf{G}^{\ell_{2}\ell_{3}\ell_{1}}_{m_{2}m_{3}m_{1}} = \iint \mathrm{d}\Omega \, Y_{\ell_{1} m_{1}}^{*}(\theta,\phi) Y_{\ell_{2}m_{2}}(\theta,\phi) Y_{\ell_{3}m_{3}}(\theta,\phi) = \sqrt{\frac{(2\ell_{2}+1)(2\ell_{3}+1)}{4\pi(2\ell_{1}+1)}}\mathsf{C}^{\ell_{2}\ell_{3}\ell_{1}}_{000}\mathsf{C}^{\ell_{2}\ell_{3}\ell_{1}}_{m_{2}m_{3}m_{1}}.
\end{equation}
\end{widetext}
The Gaunt coefficient is written in terms of the Clebsch-Gordan coefficients, $\mathsf{C}^{J_{1}J_{2}J}_{m_{1}m_{2}M}=\langle J_{1}J_{2}JM|J_{1}m_{1};J_{2}m_{2}\rangle$.  The required result for our case is 
\begin{equation}
\iint \mathrm{d}\Omega \, Y_{\ell m}^{*}(\theta,\phi) Y_{rs}^{*}(\theta,\phi) Y_{\ell'm'}(\theta,\phi) = (-1)^{s}\mathsf{G}^{r\ell'\ell}_{-sm'm}.
\end{equation}
Using the fact that $\zeta_{i}=\zeta_{LM}$, $\zeta_{j}=\zeta_{L'M'}$, and $\frac{\partial^{2}(\zeta_{\ell m}\zeta_{rs})}{\partial \zeta_{L'M'}\partial \zeta_{LM}} = \delta_{r,L}\delta_{s,M}\delta_{\ell,L'}\delta_{m,M'}+\delta_{\ell,L}\delta_{m,M}\delta_{r,L'}\delta_{s,M'}$ we find 
\begin{equation}
\kappa^{i,j} = \frac{\Delta}{2!}\sum_{\ell'm'}j_{\ell'm'}\left( (-1)^{M}\mathsf{G}^{L\ell'L'}_{-Mm'M'} + (-1)^{M'}\mathsf{G}^{L'\ell'L}_{-M'm'M} \right).
\end{equation}

The multivariate non-central moment 
\begin{equation}
\kappa^{ijk} = \langle \mathcal{C}^{i}\mathcal{C}^{j}\mathcal{C}^{k}\rangle
\end{equation}
is related to the cumulant as
\begin{equation}
\kappa^{i,j,k} = \kappa^{ijk} - \kappa^{i}\kappa^{jk} - \kappa^{j}\kappa^{ik} - \kappa^{k}\kappa^{ij} + 2 \kappa^{i}\kappa^{j}\kappa^{k}.
\end{equation}
The cumulant is computed as
\begin{equation}
\kappa^{i,j,k} = \left.\frac{\partial^{3} K_{\boldsymbol{\mathcal{C}}}(\boldsymbol{\zeta})}{\partial \zeta_{i} \partial \zeta_{j}\partial \zeta_{k}}\right|_{\boldsymbol{\zeta}=\boldsymbol{0}}.
\end{equation}
The computation requires the integral of a product of four spherical harmonics.  This is accomplished by reducing the number of spherical harmonics in the product by one, at the expense of a summation, using the relation 
\begin{equation}
Y_{rs}^{*}(\theta,\phi)Y_{pq}^{*}(\theta,\phi) = \sum_{vw}\mathsf{G}^{rpv}_{sqw}Y_{vw}^{*}(\theta,\phi).
\end{equation}
Using the fact that $\zeta_{i}=\zeta_{LM}$, $\zeta_{j}=\zeta_{L'M'}$, $\zeta_{k}=\zeta_{L''M''}$, and 
\begin{widetext}
\begin{eqnarray}
\frac{\partial^{3}(\zeta_{\ell m}\zeta_{rs}\zeta_{pq})}{\partial \zeta_{L''M''}\partial \zeta_{L'M'}\partial \zeta_{LM}} = \delta_{L,p}\delta_{M,q}\delta_{L',r}\delta_{M',s}\delta_{L'',\ell}\delta_{M'',m}+\delta_{L,p}\delta_{M,q}\delta_{L',\ell}\delta_{M',m}\delta_{L'',r}\delta_{M'',s}+\delta_{L,r}\delta_{M,s}\delta_{L',p}\delta_{M',q}\delta_{L'',\ell}\delta_{M'',m} \nonumber \\
+\delta_{L,r}\delta_{M,s}\delta_{L',\ell}\delta_{M',m}\delta_{L'',p}\delta_{M'',q}+\delta_{L,\ell}\delta_{M,m}\delta_{L',p}\delta_{M',q}\delta_{L'',r}\delta_{M'',s}+\delta_{L,\ell}\delta_{M,m}\delta_{L',r}\delta_{M',s}\delta_{L'',p}\delta_{M'',q} \nonumber \\
\end{eqnarray}
we find
\begin{eqnarray}
\kappa^{i,j,k} = \frac{\Delta}{3!}\sum_{\ell'm'}j_{\ell'm'}\sum_{vw}(-1)^{w} \left( \mathsf{G}^{L'Lv}_{M'Mw}\mathsf{G}^{\ell'vL''}_{m'-wM''} + \mathsf{G}^{L''Lv}_{M''Mw}\mathsf{G}^{\ell'vL'}_{m'-wM'} + \mathsf{G}^{LL'v}_{MM'w}\mathsf{G}^{\ell'vL''}_{m'-wM''} + \mathsf{G}^{LL''v}_{MM''w}\mathsf{G}^{\ell'vL'}_{m'-wM'} \right. \nonumber \\
\left. + \,\mathsf{G}^{L''L'v}_{M''M'w}\mathsf{G}^{\ell'vL}_{m'-wM}  +  \mathsf{G}^{L'L''v}_{M'M''w}\mathsf{G}^{\ell'vL}_{m'-wM} \right). \label{thirdcumulant}
\end{eqnarray}
\end{widetext}
Although we will not present results for any cumulants which depend on four or more indices, for example $\kappa^{i,j,k,\ell}$, it should be clear at this point that in principle there is no impediment to deriving explicit results for them if required.  The integrals to be evaluated are all products of spherical harmonics.  The Gaunt coefficients can be used to reduce products of two spherical harmonics into one, each time at the expense of a summation.  This process is repeated until the resulting product contains three spherical harmonics, and the integral can be evaluated.  As a result, the calculation of the cumulant generating function is complete, and we can move on to discuss how the results can be utilized.  
\section{Cumulants}
In this section we collect results on the cumulants up to the third order, which can be computed using the explicit equations described in the previous section.  For our purposes we first make a variable change back to the original best estimates of the concentration gradient.  Since the two variable sets in Eqs. (\ref{Cinftybest}, \ref{Cplusbest}, \ref{Cminusbest}, \ref{Czbest}) are related by constant factors, $\mathcal{C}^{i}=\beta_{i}\tilde{c}_{i}$ (no sum over $i$ here), the cumulants are easily related as
\begin{eqnarray}
\tilde{\kappa}^{i} &=& \frac{\kappa^{i}}{\beta_{i}} \label{order1} \\
\tilde{\kappa}^{i,j} &=& \frac{\kappa^{i,j}}{\beta_{i}\beta_{j}} \label{order2}\\
\tilde{\kappa}^{i,j,k} &=& \frac{\kappa^{i,j,k}}{\beta_{i}\beta_{j}\beta_{k}}.  \label{order3}
\end{eqnarray}
There is no sum over repeated indices in Eqs. (\ref{order1}, \ref{order2}, \ref{order3}).  The first order cumulants are 
\begin{eqnarray}
\tilde{\kappa}^{\infty} &=& c_{\infty} \\
\tilde{\kappa}^{+} &=& c_{+} \\
\tilde{\kappa}^{-} &=& c_{-} \\
\tilde{\kappa}^{z} &=& c_{z} 
\end{eqnarray}
The index set that was most useful from the calculational standpoint was $i \in \{\infty,+,-,z\}$.  To transform back to the Cartesian index set $i \in \{\infty,x,y,z\}$ is easy to do using the relations $c_{x}=c_{+}+c_{-}$ and $c_{y}=i(c_{-}-c_{+})$,  
\begin{eqnarray}
\tilde{\kappa}^{x} &=& \tilde{\kappa}^{+}+\tilde{\kappa}^{-} = c_{x} \\
\tilde{\kappa}^{y} &=& i(\tilde{\kappa}^{-}-\tilde{\kappa}^{+})= c_{y}. 
\end{eqnarray}

The second order cumulants are 
\begin{eqnarray}
\tilde{\kappa}^{\infty,\infty} &=& \frac{c_{\infty}}{4\pi\Delta} \\
\tilde{\kappa}^{\infty,+} &=& \frac{c_{+}}{4\pi\Delta} \\
\tilde{\kappa}^{\infty,-} &=& \frac{c_{-}}{4\pi\Delta} \\
\tilde{\kappa}^{\infty,z} &=& \frac{c_{z}}{4\pi\Delta} \\
\tilde{\kappa}^{+,-} &=& \frac{c_{\infty}}{24\pi\Delta} \\
\tilde{\kappa}^{z,z} &=& \frac{c_{\infty}}{12\pi\Delta} \\
\tilde{\kappa}^{+,+} &=& \tilde{\kappa}^{+,z} =  \tilde{\kappa}^{-,-} =\tilde{\kappa}^{-,z} = 0.
\end{eqnarray}

The Cartesian indices give 
\begin{eqnarray}
\tilde{\kappa}^{\infty,x} &=& \tilde{\kappa}^{\infty,+}+\tilde{\kappa}^{\infty,-}=\frac{c_{x}}{4\pi\Delta}\\
\tilde{\kappa}^{\infty,y} &=& i(\tilde{\kappa}^{\infty,-}-\tilde{\kappa}^{\infty,+})=\frac{c_{y}}{4\pi\Delta}\\
\tilde{\kappa}^{x,x} &=& \tilde{\kappa}^{+,+}+2\tilde{\kappa}^{+,-}+\tilde{\kappa}^{-,-}=\frac{c_{\infty}}{12\pi\Delta}\\
\tilde{\kappa}^{x,y} &=& i(\tilde{\kappa}^{-,-}-\tilde{\kappa}^{+,+})=0 \\
\tilde{\kappa}^{x,z} &=& \tilde{\kappa}^{+,z}+\tilde{\kappa}^{-,z}=0\\
\tilde{\kappa}^{y,y} &=& -\tilde{\kappa}^{+,+}+2\tilde{\kappa}^{+,-}-\tilde{\kappa}^{-,-}=\frac{c_{\infty}}{12\pi\Delta}\\
\tilde{\kappa}^{y,z} &=& i(\tilde{\kappa}^{-,z}-\tilde{\kappa}^{+,z}) = 0.
\end{eqnarray}

The third order cumulants are 
\begin{eqnarray}
\tilde{\kappa}^{\infty,\infty,\infty} &=& \frac{c_{\infty}}{16\pi^{2}\Delta^{2}}\\
\tilde{\kappa}^{\infty,\infty,+} &=& \frac{c_{+}}{16\pi^{2}\Delta^{2}}\\
\tilde{\kappa}^{\infty,\infty,-} &=& \frac{c_{-}}{16\pi^{2}\Delta^{2}}\\
\tilde{\kappa}^{\infty,\infty,z} &=& \frac{c_{z}}{16\pi^{2}\Delta^{2}}\\
\tilde{\kappa}^{\infty,+,-} &=& \frac{c_{\infty}}{96\pi^{2}\Delta^{2}}\\
\tilde{\kappa}^{\infty,z,z} &=& \frac{c_{\infty}}{48\pi^{2}\Delta^{2}}\\
\tilde{\kappa}^{+,+,-} &=& \frac{c_{+}}{80\pi^2\Delta^2} \\
\tilde{\kappa}^{+,-,-} &=& \frac{c_{-}}{80\pi^2\Delta^2} \\
\tilde{\kappa}^{+,-,z} &=& \frac{c_{z}}{160\pi^2\Delta^2} \\
\tilde{\kappa}^{+,z,z} &=& \frac{c_{+}}{80\pi^2\Delta^2} \\
\tilde{\kappa}^{-,z,z} &=& \frac{c_{-}}{80\pi^2\Delta^2} \\
\tilde{\kappa}^{z,z,z} &=& \frac{3c_{z}}{80\pi^2\Delta^2} \\
\tilde{\kappa}^{\infty,+,+} &=& \tilde{\kappa}^{\infty,+,z} = \tilde{\kappa}^{\infty,-,-} =  \tilde{\kappa}^{\infty,-,z} =0 \\
\tilde{\kappa}^{+,+,+} &=& \tilde{\kappa}^{+,+,z}= \tilde{\kappa}^{-,-,-}=\tilde{\kappa}^{-,-,z}=0
\end{eqnarray}

The Cartesian indices give
\begin{eqnarray}
\tilde{\kappa}^{\infty,\infty,x} &=& \tilde{\kappa}^{\infty,\infty,+} + \tilde{\kappa}^{\infty,\infty,-} = \frac{c_{x}}{16\pi^{2}\Delta^{2}} \\
\tilde{\kappa}^{\infty,\infty,y} &=& i( \tilde{\kappa}^{\infty,\infty,-} - \tilde{\kappa}^{\infty,\infty,+}) = \frac{c_{y}}{16\pi^{2}\Delta^{2}} \\
\tilde{\kappa}^{\infty,x,y} &=& i(  \tilde{\kappa}^{\infty,-,-} -\tilde{\kappa}^{\infty,+,+}  ) = 0 \\
\tilde{\kappa}^{x,x,y} &=& i( \tilde{\kappa}^{+,-,-} - \tilde{\kappa}^{+,+,-} ) = \frac{c_{y}}{80\pi^{2}\Delta^{2}} \\
\tilde{\kappa}^{x,y,y} &=& \tilde{\kappa}^{+,+,-} + \tilde{\kappa}^{+,-,-} = \frac{c_{x}}{80\pi^{2}\Delta^{2}} \\
\tilde{\kappa}^{x,y,z} &=& i(\tilde{\kappa}^{-,-,z}-\tilde{\kappa}^{+,+,z})=0 \\
\tilde{\kappa}^{x,z,z} &=& \tilde{\kappa}^{+,z,z} + \tilde{\kappa}^{-,z,z} = \frac{c_{x}}{80\pi^{2}\Delta^{2}}\\
\tilde{\kappa}^{y,z,z} &=& i(\tilde{\kappa}^{-,z,z} - \tilde{\kappa}^{+,z,z}) = \frac{c_{y}}{80\pi^{2}\Delta^{2}}\\
\tilde{\kappa}^{\infty,x,x} &=& 2\tilde{\kappa}^{\infty,+,-} = \frac{c_{\infty}}{48\pi^{2}\Delta^{2}}\\
\tilde{\kappa}^{\infty,x,z} &=& \tilde{\kappa}^{\infty,+,z} + \tilde{\kappa}^{\infty,-,z}  = 0\\
\tilde{\kappa}^{\infty,y,y} &=& 2\tilde{\kappa}^{\infty,+,-} = \frac{c_{\infty}}{48\pi^{2}\Delta^{2}}\\
\tilde{\kappa}^{\infty,y,z} &=& i(\tilde{\kappa}^{\infty,-,z} - \tilde{\kappa}^{\infty,+,z})  = 0\\
\tilde{\kappa}^{x,x,x} &=& 3(\tilde{\kappa}^{+,+,-} + \tilde{\kappa}^{+,-,-})  =  \frac{3c_{x}}{80\pi^2\Delta^2} \\
\tilde{\kappa}^{x,x,z} &=& 2\tilde{\kappa}^{+,-,z} =   \frac{c_{z}}{80\pi^2\Delta^2}\\
\tilde{\kappa}^{y,y,y} &=& 3i(\tilde{\kappa}^{+,-,-}-\tilde{\kappa}^{+,+,-}) =  \frac{3c_{y}}{80\pi^2\Delta^2} \\
\tilde{\kappa}^{y,y,z} &=& 2\tilde{\kappa}^{+,-,z} =   \frac{c_{z}}{80\pi^2\Delta^2}
\end{eqnarray}

\section{Chemotactic Index}

With these results in hand, we turn our attention to making connection to the chemotaxis experiments.  In the experiments, the chemotactic index $\Psi$ was measured for {\it Dictyostelium discoideum} cells in a cAMP gradient.  The chemotactic index is defined as the distance travelled by cells in the direction of the true gradient divided by the total distance moved.  Endres and Wingreen computed an {\it optimal} $\Psi$ making the following assumptions.  After averaging for a time $T$ the cell moves with constant velocity in the direction of the estimated gradient.  Considering two-dimensional motion in the $x$-$z$ plane, we are free to choose our coordinate system so that the true gradient points in the $z$ direction.  Assuming the same constant velocity and run time for each run, the run lengths $l$ will also be constant.  The chemotactic index for run $i$ is $\cos \theta_{i}$, where $\theta_{i}$ is the angle between the true gradient and the estimated gradient.  Then the average chemotactic index
\begin{equation}
\Psi = \frac{\sum_{i=1}^{N}l_{i,z}}{\sum_{i=1}^{N}l_{i}} = \frac{l\sum_{i=1}^{N}\cos\theta_{i}}{Nl} = \mathbb{E}[\cos \theta].
\end{equation}
Using our best estimates of the gradient we can write $\cos \theta = \frac{\tilde{c}_{z}}{\sqrt{\tilde{c}_{x}^{2}+\tilde{c}_{z}^{2}}}$.  The goal is then to calculate the expectation value $\mathbb{E}\left[ \frac{\tilde{c}_{z}}{\sqrt{\tilde{c}_{x}^{2}+\tilde{c}_{z}^{2}}}\right]$.  

We are led naturally to the following inverse problem.  From our knowledge of the cumulants, estimate the unknown joint distribution $P(\boldsymbol{\tilde{c}})=P(\tilde{c}_{\infty},\tilde{c}_{x},\tilde{c}_{y},\tilde{c}_{z})$.  Marginalize over $\tilde{c}_{\infty}$ and $\tilde{c}_{y}$, obtaining 
\begin{equation}
P(\tilde{c}_{x},\tilde{c}_{z}) = \iint P(\tilde{c}_{\infty},\tilde{c}_{x},\tilde{c}_{y},\tilde{c}_{z}) \mathrm{d}\tilde{c}_{\infty}\mathrm{d}\tilde{c}_{y}. 
\end{equation}
Use this marginal distribution to compute the chemotactic index, 
\begin{equation}
\Psi = \iint P(\tilde{c}_{x},\tilde{c}_{z}) \frac{\tilde{c}_{z}}{\sqrt{\tilde{c}_{x}^{2}+\tilde{c}_{z}^{2}}}  \mathrm{d}\tilde{c}_{x}\mathrm{d}\tilde{c}_{z}.
\end{equation}
In \cite{endres2008accuracy} there is not a discussion of this procedure in the paper or supplementary material, as the starting point is an {\it assumption } of a marginal distribution of the Gaussian form,
\begin{equation}
P(\tilde{c}_{x},\tilde{c}_{z})=\frac{1}{2\pi\sigma _{x}\sigma_{z}} e^{-\frac{1}{2}\left(\frac{\tilde{c}_{x}}{\sigma_{x}}\right)^{2}-\frac{1}{2}\left(\frac{\tilde{c}_{z}-c_{z}}{\sigma_{z}}\right)^{2}}.   \label{marginal}
\end{equation}
Here $\sigma_{x}^{2}=\tilde{\kappa}^{x,x}$, $\sigma_{z}^{2}=\tilde{\kappa}^{z,z}$, and we have used the fact that since the true gradient is in the $z$ direction, $\tilde{\kappa}^{x}=c_{x}=0$.  In this case the result of the integration is 
\begin{eqnarray}
\Psi &=& \left(\frac{\pi s}{2}\right)^{\frac{1}{2}}e^{-s}(I_{0}(s)+I_{1}(s))  \label{Psigauss} \\
s &=& \left(\frac{c_{z}}{2\sigma_{z}}\right)^{2} = \frac{3\pi c_{z}^{2}\Delta}{c_{\infty}}.  \label{signoise}
\end{eqnarray}
Here $I_{0}(s)$ and $I_{1}(s)$ are the modified Bessel functions of the first kind of order 0 and 1, respectively.  The chemotactic index depends on the dimensionless groups in the problem only through the combination $s$, which can be interpreted as the square of the signal to noise ratio.  The signal is proportional to the true gradient $c_{z}$, and $2\sigma_{z}$ is a measure of the noise due to particle diffusion.  

One might wonder under what circumstances this calculation can be justified.  If the result is not exact, what is the nature of the approximation scheme?  A joint distribution which is Gaussian,
\begin{equation}
P(\boldsymbol{\tilde{c}}) = \mathcal{G}(\boldsymbol{\tilde{c}}) = \frac{e^{-\frac{1}{2}(\tilde{c}^{i}-\tilde{\kappa}^{i})\tilde{\kappa}_{i,j}(\tilde{c}^{j}-\tilde{\kappa}^{j})}}{\sqrt{(2\pi)^{4}\det(\tilde{\kappa}^{(2)})}},
\end{equation}
will have the marginal distribution Eq. (\ref{marginal}).  Here $\tilde{c}^{i}$ is the $i^{th}$ component of the concentration vector $\boldsymbol{\tilde{c}}=(\tilde{c}_{\infty},\tilde{c}_{x},\tilde{c}_{y},\tilde{c}_{z})$ using the Cartesian index set $i \in \{\infty, x, y, z\}$, $\tilde{\kappa}_{i,j}=(\tilde{\kappa}^{-1})^{i,j}$ are the components of the inverse of the covariance matrix, and 
\begin{equation}
\det(\tilde{\kappa}^{(2)}) = \sum_{i_{1},i_{2},i_{3},i_{4}} \varepsilon_{i_{1}i_{2}i_{3}i_{4}}\tilde{\kappa}^{\infty,i_{1}}\tilde{\kappa}^{x,i_{2}}\tilde{\kappa}^{y,i_{3}}\tilde{\kappa}^{z,i_{4}}
\end{equation}
is the determinant of the covariance matrix.  Here $\varepsilon_{i_{1}i_{2}i_{3}i_{4}}$ is the Levi-Civita symbol.  

Intuitively, one might hope that the cumulants $\tilde{\kappa}^{i_{1},i_{2},\cdots,i_{n}}$ will be negligible if $n$ is sufficiently large, and can be neglected beyond some specified order.  However, according to a multivariate  generalization of the  Marcinkiewicz 
theorem \cite{rajagopal1974some}, if the cumulant generating function of a probability distribution in $k$ variables is a multinomial of degree greater than 2, then the probability distribution will not be positive definite.  So, there is not a 
good justification for truncation of the cumulant generating function, Eq. (\ref{cumulantgenerating2}).  The Gaussian distribution is, in fact, special, as it is the only distribution with zero cumulants at all orders greater than 2.  In our case we have computed the non-zero cumulants of order 3, which means there are also non-zero cumulants at any order, and the joint distribution is {\it not} a Gaussian.  The calculation of the chemotactic index needs to be modified to capture the non-Gaussianity of the distribution.  

\section{Edgeworth Expansion}

To move beyond this approximation, we can use the higher order cumulants to approximate the non-Gaussian joint distribution as an asymptotic series, referred to as an Edgeworth expansion \cite{mccullagh2018tensor,sellentin2017use}.  The first term in the expansion yields a joint distribution of the form 
\begin{equation}
P(\boldsymbol{\tilde{c}}) \approx  \mathcal{G}(\boldsymbol{\tilde{c}}) \left( 1 + \frac{\tilde{\kappa}^{i,j,k}}{3!}h_{ijk}\right) \label{edgeworth}
\end{equation}
in terms of the multivariate Hermite tensors, 
\begin{eqnarray}
h_{i} &=& \tilde{\kappa}_{i,j}(\tilde{c}^{j}-\tilde{\kappa}^{j})\\
h_{ij} &=& h_{i}h_{j} - \tilde{\kappa}_{i,j} \\
h_{ijk} &=& h_{i}h_{j}h_{k} - h_{i}\tilde{\kappa}_{j,k} -  h_{j}\tilde{\kappa}_{i,k} -  h_{k}\tilde{\kappa}_{i,j}.
\end{eqnarray}
The marginal distribution is no longer a Gaussian, but can be expressed as 
\begin{equation}
P(\tilde{c}_{x},\tilde{c}_{z})= e^{-\frac{1}{2}\left(\frac{\tilde{c}_{x}}{\sigma_{x}}\right)^{2}-\frac{1}{2}\left(\frac{\tilde{c}_{z}-c_{z}}{\sigma_{z}}\right)^{2}}\sum_{m=0}^{3}\sum_{n=0}^{3} a_{mn}(\tilde{c}_{x})^{m}(\tilde{c}_{z}-c_{z})^{n}. \label{margedgeworth}
\end{equation}
The derivation is given in Appendix A, along with the values for all of the coefficients $a_{mn}$.  As a result, we compute a new result for the chemotactic index with this modified marginal distribution.  The result depends on four terms from the expansion, and can be expressed as 
\begin{equation}
\Psi = \Psi_{00}+\Psi_{01}+\Psi_{21}+\Psi_{03}.  
\end{equation}
The first term in the expansion, $\Psi_{00}$, is identical to the result from the Gaussian approximation, Eq. (\ref{Psigauss}).  All terms can be calculated as follows, see Appendix B for the derivation, 
\begin{widetext}
\begin{eqnarray}
\Psi_{mn}&=&\left( \frac{\sigma_{z}^{2}}{c_{z}}\right)^{2+m+n}e^{-2s}a_{mn}\mathcal{R}_{mn}(s) \\
\mathcal{R}_{mn}(s) &=&  \int_{0}^{\infty}   \alpha^{1+m+n}e^{-\frac{\alpha^{2}}{8s}}\mathcal{I}_{mn}(\alpha,s) \mathrm{d}\alpha \\
\mathcal{I}_{mn}(\alpha,s) &=& \int_{0}^{2\pi}e^{\alpha \cos \theta} (\sin\theta)^{m} \left(\cos \theta - \frac{4s}{\alpha} \right)^{n} \cos \theta \,\mathrm{d}\theta.
\end{eqnarray}
\end{widetext}

\section{Connection to Experiments}

The theoretical calculation of the chemotactic index can be compared to the experimental observations \cite{VANHAASTERT20071787}.  In the original experiments, there were three different experimental assays studied, which were called the micropipette assay, the modified Zigmond chamber assay, and the small population assay.  The small population assay involves measuring the response of cells to chemoattractant (cAMP) applied as a single dose.  As a result, a transient cAMP gradient is established, but there is never a steady state concentration.  It is questionable that the model development, which pertains to a steady-state concentration gradient, is even applicable in this case.  As a result, we do not consider comparison to the small population assay in this paper.  

The modified Zigmond chamber assay is potentially the simplest system to compare to the theoretical model.  The reason is that it establishes an experimental steady state concentration gradient of the form (see Appendix A in \cite{VANHAASTERT20071787})
\begin{equation}
C(Z) = C_{s}\left(1-\frac{Z}{L}\right),
\end{equation}
where $C_{s}$ is the cAMP concentration at the source and $L$ is the width of the bridge (2000 $\mu m$ in the experiments).  Here $Z$ is the distance from the cell to the source.  This is the same form as the far-field concentration profile in the theoretical model, and as a result we can immediately compute the dimensionless groups of the model.  We find $c_{\infty}=C_{s}R^{3}$ and $c_{z}=-\frac{C_{s}R^{4}}{L}$, where $R=5 \,\mu m$ is the cell radius.  Importantly, since the parameters are constants, all cells in the field of view would experience the same gradient profile.  The experiments investigated source concentrations of 10 $nM$, 100 $nM$, and 1000 $nM$.  We find that $s\sim 1.7$, $s \sim 17$, and $s\sim 170$ in the three cases, respectively.  The importance of corrections to the Gaussian approximation for the chemotactic index is quantified by the dimensionless parameter $\lambda=\frac{c_{z}}{c_{\infty}}$, with corrections $\mathcal{O}(\lambda)$.  Since $\lambda=-\frac{R}{L}= -0.0025$ is independent of the source concentration, in all cases we have $|\lambda|  \ll  s$.  We refer to this order of magnitude separation between the dimensionless groups of the problem as experimental cases in which the gradients are shallow.  As a result, the Gaussian approximation is sufficient to describe these experimental results.  

In the micropipette assay, the experimental steady state concentration profile is well described by 
\begin{equation}
C(Z)=\frac{\alpha C_{p}}{Z}.  
\end{equation}
Here $C_{p}$ is the cAMP concentration in the pipette and $\alpha=0.05 \, \mu m$ is an experimentally determined length scale that depends on the geometry of the pipette and the applied pressure.  In this case, with $z=Z/R$ the dimensionless distance of the cell from the pipette, we find that the dimensionless groups of the problem are position dependent, 
\begin{eqnarray}
c_{\infty} &=& \frac{\gamma}{z} \\
c_{z} &=& -\frac{\gamma}{z^{2}},
\end{eqnarray}
where $\gamma = \alpha R^{2} C_{p}$ is a dimensionless experimental constant.  Cells at different locations experience different concentration gradients.  Hence, the square of the signal to noise ratio inherits this distance dependence, as 
\begin{equation}
s = \frac{3\pi \gamma \Delta}{z^{3}}.  
\end{equation}
To move beyond the Gaussian approximation we also need to compute 
\begin{equation}
\lambda = -\frac{1}{z}.
\end{equation}
\\
With a cAMP diffusion coefficient $D=300 \, \mu m^{2}s^{-1}$, an averaging time of $T=3.2\, s$, and $R=5\, \mu m$ we compute the dimensionless averaging time $\Delta=38.4$.  Experiments explored pipette concentrations of 0.1 $\mu M$, 1 $\mu M$, 10 $\mu M$, and 100 $\mu M$.  Because $\gamma$ is proportional to $C_{p}$, the corrections will be most relevant at the smallest pipette concentration of $C_{p}=0.1\, \mu M$, where we find $\gamma=75.3$.  
\begin{figure}[h]
\centering{\includegraphics[width=20pc]{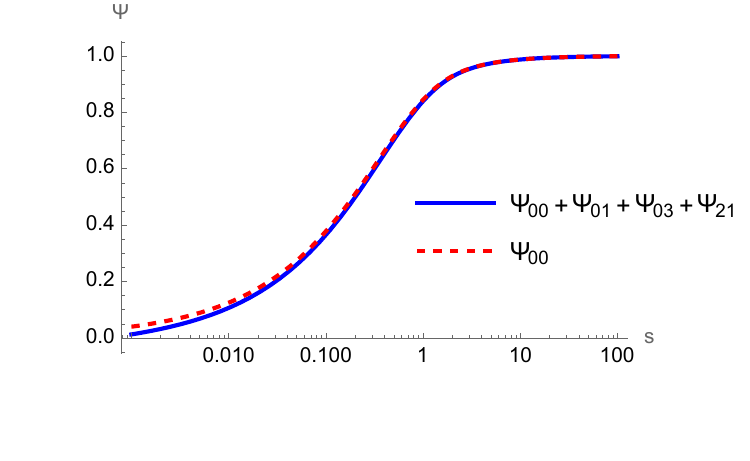}}
\caption{{\bf Chemotactic Index $\Psi$ in the micropipette assay.} The Gaussian approximation $\Psi_{00}$ plotted as a function of $s$, and the Edgeworth approximation $\Psi_{00}+\Psi_{01}+\Psi_{21}+\Psi_{03}$ with $C_{p}=0.1 \, \mu M$.  }
\label{psiplot}
\end{figure}
Even so, the corrections are again very small in this case as demonstrated in Fig. \ref{psiplot}.  One can compare our Fig. \ref{psiplot} to the inset to Fig. 1 in \cite{endres2008accuracy}.  The results accurately describe the experiments, apart from one issue.  In the experiments, the chemotactic index saturates at approximately 0.9 at large $s$ ($z \sim 1$), whereas the theoretical predictions saturate at 1.  As a result, in Fig. 1 in \cite{endres2008accuracy}, the theoretical curves were rescaled by this factor of 0.9 to match the experiments.   

\begin{figure}[h]
\centering{\includegraphics[width=20pc]{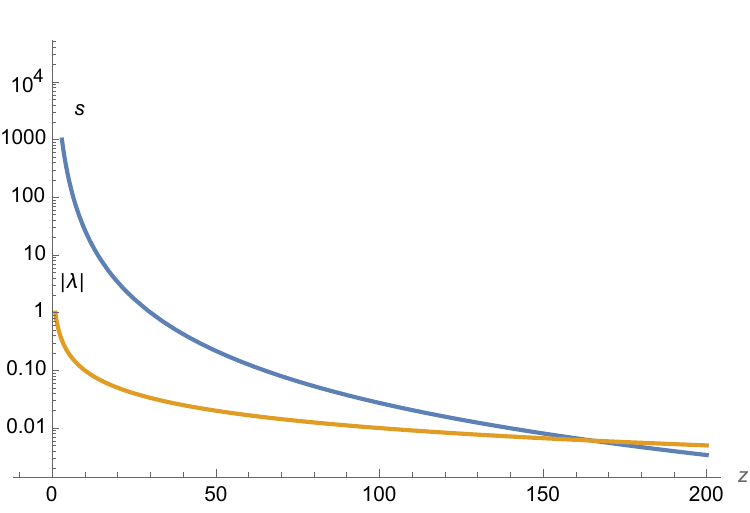}}
\caption{{\bf Dimensionless groups in the micropipette assay.} The square of the signal to noise ratio $s$ and the magnitude of the concentration ratio $|\lambda|$ plotted as a function of the dimensionless distance of the cell from the pipette $z$ in the micropipette assay with $C_{p}=0.1 \, \mu M$.  }
\label{dimparams}
\end{figure}

The smallness of the corrections here is not a generic result, but rather a consequence of the fact that in the experimental system studied, it is only in the region far from the pipette ($z \gg 1$) when the square of the signal to noise ratio is small ($s \ll 1$) that the corrections become relevant ($ |\lambda|\sim s$).  Note that for $s \ll 1$ we have $\Psi_{00} \sim \mathcal{O}(s^{\frac{1}{2}})$ whereas corrections to the chemotactic index are $\Psi_{01},\Psi_{21},\Psi_{03} \sim \mathcal{O}(\lambda s^{-\frac{1}{2}})$.  In Fig. \ref{dimparams} we investigate the magnitude of the dimensionless groups in the problem, demonstrating this behavior.  The micropipette assay experiments are all in the regime of shallow gradients.  

\section{Conclusion}

In this paper we theoretically investigated the chemotactic index $\Psi$ for single cells responding to a diffusing chemoattractant.  The error minimization formalism is a regression problem (Eq. \ref{dimE}) which can be solved exactly for the cumulants in the model.  The original model formulation focused only on the mean and variance, but here we demonstrate that the multivariate cumulant generating function has a closed form expression in terms of products of spherical harmonics (Eq. \ref{cumulantgenerating}).  We derived explicit results for the third order cumulants (Eq. \ref{thirdcumulant}), which allowed us to compute the leading terms in an Edgeworth expansion for the non-Gaussian marginal distribution required to compute the chemotactic index (Eq. \ref{margedgeworth}).  We identified a dimensionless concentration ratio $\lambda$ (Eq. \ref{lambda}) which captures the effects of the non-Gaussianity, and derived explicit expressions for the leading corrections to the chemotactic index (Eqs. \ref{psiresult2}, \ref{psiresult3}, \ref{psiresult4}).  The original Gaussian approximation for the chemotactic index (Eq. \ref{psiresult1}), which depends only on the square of the signal to noise ratio $s$, is an excellent approximation in the experimental systems studied.  We demonstrated that the close agreement between the experiments and the Gaussian theory relies crucially on the fact that in the experimental systems that were investigated, the concentration gradients were shallow, $|\lambda| \ll s$.  Future synthetic experiments, as well as the diverse natural environments encountered by cells, may probe regimes in parameter space where gradients are not shallow, and these corrections become relevant for correctly interpreting cells' chemotactic efficiency.  Although the specific results for the chemotactic index in this paper are limited to planar cell migration in two spatial dimensions, the formalism could be extended to treat the case of cell migration in three spatial dimensions.  This is a task for future research.  

\appendix

\begin{widetext}

\section{Marginal Distribution from the Edgeworth Expansion}

We write the result for the joint distribution, Eq. (\ref{edgeworth}), in the following form, where $\mathcal{N}$ is a normalization factor to be determined.  Isolating the dependence on $\tilde{c}_{y}$ we have
\begin{equation}
P(\boldsymbol{\tilde{c}})=\mathcal{N}e^{Q_{0}-Q_{2}\tilde{c}_{y}^{2}}(A_{0}+A_{1}\tilde{c}_{y}+A_{2}\tilde{c}_{y}^{2}+A_{3}\tilde{c}_{y}^{3})
\end{equation}
where $Q_{2}=\frac{6\pi\Delta}{c_{\infty}}$.  Here $Q_{0}$, $A_{0}$, $A_{1}$, $A_{2}$, and $A_{3}$ depend on the remaining variables $\tilde{c}_{\infty}$, $\tilde{c}_{x}$, and $\tilde{c}_{z}$.  Integrating over $\tilde{c}_{y}$ the odd terms vanish by symmetry.  Using the results of the Gaussian integrals
\begin{eqnarray}
\int_{-\infty}^{\infty} e^{-Q_{2}\tilde{c}_{y}^{2}} \, \mathrm{d}\tilde{c}_{y} &=& \left(\frac{\pi}{Q_{2}}\right)^{\frac{1}{2}} \\
\int_{-\infty}^{\infty} \tilde{c}_{y}^{2} \,e^{-Q_{2}\tilde{c}_{y}^{2}} \, \mathrm{d}\tilde{c}_{y} &=& \frac{\pi^{\frac{1}{2}}}{2(Q_{2})^{\frac{3}{2}}}
\end{eqnarray}
we have 
\begin{equation}
\int_{-\infty}^{\infty} P(\boldsymbol{\tilde{c}})\mathrm{d}\tilde{c}_{y} = \mathcal{N}e^{Q_{0}}\left( A_{0} \left(\frac{\pi}{Q_{2}}\right)^{\frac{1}{2}} + A_{2} \frac{\pi^{\frac{1}{2}}}{2(Q_{2})^{\frac{3}{2}}} \right).
\end{equation}
Now isolating the dependence on $\tilde{c}_{\infty}$ we have
\begin{eqnarray}
A_{0} &=& B_{0} + B_{1} \tilde{c}_{\infty} + B_{2} \tilde{c}_{\infty}^{2} + B_{3} \tilde{c}_{\infty}^{3} \\
A_{2} &=& D_{0} + D_{1}\tilde{c}_{\infty} \\
Q_{0} &=& \psi_{0} + \psi_{1}\tilde{c}_{\infty} - \psi _{2}\tilde{c}_{\infty}^{2}.
\end{eqnarray}
Here $B_{0}$, $B_{1}$, $B_{2}$, $B_{3}$, $D_{0}$, $D_{1}$, $\psi_{0}$, $\psi_{1}$, and $\psi_{2}$ depend on the variables $\tilde{c}_{x}$ and $\tilde{c}_{z}$.  We make a tiny error extending the range of the integration over $\tilde{c}_{\infty}$ from $[0,\infty)$ to $(-\infty,\infty)$.  Using the results for the Gaussian integrals
\begin{eqnarray}
\int_{-\infty}^{\infty} e^{-\psi_{2}\tilde{c}_{\infty}^{2}+\psi_{1}\tilde{c}_{\infty}} \, \mathrm{d}\tilde{c}_{\infty} &=& e^{\frac{(\psi_{1})^{2}}{4\psi_{2}}}\left(\frac{\pi}{\psi_{2}}\right)^{\frac{1}{2}} \\
\int_{-\infty}^{\infty} \tilde{c}_{\infty}e^{-\psi_{2}\tilde{c}_{\infty}^{2}+\psi_{1}\tilde{c}_{\infty}} \, \mathrm{d}\tilde{c}_{\infty} &=& e^{\frac{(\psi_{1})^{2}}{4\psi_{2}}} \frac{\psi_{1}\pi^{\frac{1}{2}}}{2(\psi_{2})^{\frac{3}{2}}} \\
\int_{-\infty}^{\infty} \tilde{c}_{\infty}^{2}e^{-\psi_{2}\tilde{c}_{\infty}^{2}+\psi_{1}\tilde{c}_{\infty}} \, \mathrm{d}\tilde{c}_{\infty} &=& e^{\frac{(\psi_{1})^{2}}{4\psi_{2}}} \frac{(2\psi_{2}+(\psi_{1})^{2})\pi^{\frac{1}{2}}}{4(\psi_{2})^{\frac{5}{2}}} \\
\int_{-\infty}^{\infty} \tilde{c}_{\infty}^{3}e^{-\psi_{2}\tilde{c}_{\infty}^{2}+\psi_{1}\tilde{c}_{\infty}} \, \mathrm{d}\tilde{c}_{\infty} &=& e^{\frac{(\psi_{1})^{2}}{4\psi_{2}}} \frac{\psi_{1}(6\psi_{2}+(\psi_{1})^{2})\pi^{\frac{1}{2}}}{8(\psi_{2})^{\frac{7}{2}}} 
\end{eqnarray}
we find
\begin{eqnarray}
\int_{-\infty}^{\infty} \int_{-\infty}^{\infty}  P(\boldsymbol{\tilde{c}})\mathrm{d}\tilde{c}_{y} \mathrm{d}\tilde{c}_{\infty} =\mathcal{N}e^{\psi_{0}+\frac{(\psi_{1})^{2}}{4\psi_{2}}} \left[ \left(B_{0}\left(\frac{\pi}{Q_{2}}\right)^{\frac{1}{2}} + D_{0}\frac{\pi^{\frac{1}{2}}}{2(Q_{2})^{\frac{3}{2}}}\right)\left(\frac{\pi}{\psi_{2}}\right)^{\frac{1}{2}} + \right. \nonumber \\
\left. \left(B_{1}\left(\frac{\pi}{Q_{2}}\right)^{\frac{1}{2}} + D_{1}\frac{\pi^{\frac{1}{2}}}{2(Q_{2})^{\frac{3}{2}}}\right) \frac{\psi_{1}\pi^{\frac{1}{2}}}{2(\psi_{2})^{\frac{3}{2}}} + B_{2}\left(\frac{\pi}{Q_{2}}\right)^{\frac{1}{2}}\frac{(2\psi_{2}+(\psi_{1})^{2})\pi^{\frac{1}{2}}}{4(\psi_{2})^{\frac{5}{2}}} +  B_{3}\left(\frac{\pi}{Q_{2}}\right)^{\frac{1}{2}}\frac{\psi_{1}(6\psi_{2}+(\psi_{1})^{2})\pi^{\frac{1}{2}}}{8(\psi_{2})^{\frac{7}{2}}}\right].
\end{eqnarray}
Computing the argument of the exponent, $\psi_{0}+\frac{(\psi_{1})^{2}}{4\psi_{2}}=Q_{2}$, and isolating the dependence on $\tilde{c}_{x}$, this result can be written in the form
\begin{equation}
\int_{-\infty}^{\infty} \int_{-\infty}^{\infty}  P(\boldsymbol{\tilde{c}})\mathrm{d}\tilde{c}_{y} \mathrm{d}\tilde{c}_{\infty} =\mathcal{N}e^{-Q_{2}\tilde{c}_{x}^{2}-Q_{2}\left(\tilde{c}_{z}-c_{z}\right)^{2}}[F_{0}+F_{1}\tilde{c}_{x}+F_{2}\tilde{c}_{x}^{2}+F_{3}\tilde{c}_{x}^{3}]. \label{margcx}
\end{equation}
Here $F_{0}$, $F_{1}$, $F_{2}$, and $F_{3}$ depend on $\tilde{c}_{z}$.  Performing the integration over $\tilde{c}_{x}$, again the odd terms drop out by symmetry, and we find 
\begin{equation}
\int_{-\infty}^{\infty} \int_{-\infty}^{\infty} \int_{-\infty}^{\infty}  P(\boldsymbol{\tilde{c}})\mathrm{d}\tilde{c}_{y} \mathrm{d}\tilde{c}_{\infty} \mathrm{d}\tilde{c}_{x} =\mathcal{N}e^{-Q_{2}\left(\tilde{c}_{z}-c_{z}\right)^{2}}\left[F_{0} \left(\frac{\pi}{Q_{2}}\right)^{\frac{1}{2}}+F_{2} \frac{\pi^{\frac{1}{2}}}{2(Q_{2})^{\frac{3}{2}}}\right].
\end{equation}
This result can be written in the form
\begin{equation}
\int_{-\infty}^{\infty} \int_{-\infty}^{\infty} \int_{-\infty}^{\infty}  P(\boldsymbol{\tilde{c}})\mathrm{d}\tilde{c}_{y} \mathrm{d}\tilde{c}_{\infty} \mathrm{d}\tilde{c}_{x} =\mathcal{N}e^{-Q_{2}\left(\tilde{c}_{z}-c_{z}\right)^{2}} \left[ M_{0} + M_{1}(\tilde{c}_{z}-c_{z}) + M_{3} (\tilde{c}_{z}-c_{z})^{3} \right].
\end{equation}
Performing the remaining integration over $\tilde{c}_{z}$, the linear and cubic term vanish by symmetry, leaving
\begin{equation}
\iiiint P(\boldsymbol{\tilde{c}}) \mathrm{d}^{4}\boldsymbol{\tilde{c}} = \int_{-\infty}^{\infty} \int_{-\infty}^{\infty} \int_{-\infty}^{\infty} \int_{-\infty}^{\infty}  P(\boldsymbol{\tilde{c}})\mathrm{d}\tilde{c}_{y} \mathrm{d}\tilde{c}_{\infty} \mathrm{d}\tilde{c}_{x}  \mathrm{d}\tilde{c}_{z} =\mathcal{N} M_{0} \left(\frac{\pi}{Q_{2}}\right)^{\frac{1}{2}}.
\end{equation}
Requiring that the joint distribution is normalized we find 
\begin{equation}
\mathcal{N} = \left(\frac{Q_{2}}{\pi}\right)^{\frac{1}{2}}\frac{1}{M_{0}} = \frac{12\sqrt{3}\Delta^{2}}{c_{\infty}(c_{\infty}^{2}-3c_{z}^{2})^{\frac{1}{2}}}.
\end{equation}
Collecting these results, the marginal distribution Eq. (\ref{margcx}) can be written as an exponential multiplied by a multinomial, in the following form
\begin{equation}
P(\tilde{c}_{x},\tilde{c}_{z})=\int_{-\infty}^{\infty} \int_{-\infty}^{\infty}  P(\boldsymbol{\tilde{c}})\mathrm{d}\tilde{c}_{y} \mathrm{d}\tilde{c}_{\infty} = e^{-\frac{1}{2}\left(\frac{\tilde{c}_{x}}{\sigma_{x}}\right)^{2}-\frac{1}{2}\left(\frac{\tilde{c}_{z}-c_{z}}{\sigma_{z}}\right)^{2}}\sum_{m=0}^{3}\sum_{n=0}^{3} a_{mn}(\tilde{c}_{x})^{m}(\tilde{c}_{z}-c_{z})^{n}.
\end{equation}
The coefficients $a_{mn}$ are given by
\begin{eqnarray}
a_{00} &=& \frac{6\Delta}{c_{\infty}} \\
a_{10} &=& \frac{4\Delta}{5}\left( \frac{1}{c_{\infty}^{2}-3c_{z}^{2}}-\frac{(6c_{\infty}+9c_{z})}{c_{\infty}^{3}}\right)\\
a_{20} &=& 0 \\
a_{30} &=& \frac{36\pi\Delta^{2}(5c_{\infty}+3c_{z})}{5c_{\infty}^{4}} \\
a_{01} &=& -\frac{2\Delta(10c_{\infty}^{3}+54c_{\infty}^{2}c_{z}+45c_{\infty}c_{z}^{2}-27c_{z}^{3})}{5c_{\infty}^{3}(c_{\infty}^{2}-3c_{z}^{2})} \\
a_{11} &=& 0 \\
a_{21} &=& \frac{36\pi\Delta^{2}(5c_{\infty}^{3}+14c_{\infty}^{2}c_{z}+3c_{\infty}c_{z}^{2}-18c_{z}^{3})}{5c_{\infty}^{4}(c_{\infty}^{2}-3c_{z}^{2})} \\
a_{31} &=& 0 \\
a_{02} &=& 0 \\
a_{12} &=& \frac{36\pi\Delta^{2}(c_{\infty}+3c_{z})(5c_{\infty}^{2}-9c_{z}^{2})}{5c_{\infty}^{4}(c_{\infty}^{2}-3c_{z}^{2})} \\
a_{22} &=& 0 \\
a_{32} &=& 0 \\
a_{03} &=& \frac{36\pi\Delta^{2}(c_{\infty}+c_{z})(5c_{\infty}+9c_{z})}{5c_{\infty}^{3}(c_{\infty}^{2}-3c_{z}^{2})} \\
a_{13} &=& 0 \\
a_{23} &=& 0 \\
a_{33} &=& 0.
\end{eqnarray}

\section{Computation of the Chemotactic Index $\Psi$}

We begin with the definition of the chemotactic index, 
\begin{equation}
\Psi = \mathbb{E}\left[ \frac{\tilde{c}_{z}}{\sqrt{\tilde{c}_{x}^{2}+\tilde{c}_{z}^{2}}}\right] = \iint P(\tilde{c}_{x},\tilde{c}_{z}) \frac{\tilde{c}_{z}}{\sqrt{\tilde{c}_{x}^{2}+\tilde{c}_{z}^{2}}}  \mathrm{d}\tilde{c}_{x}\mathrm{d}\tilde{c}_{z}.
\end{equation}
Using the result for the marginal distribution, Eq. (\ref{margedgeworth}), we have
\begin{equation}
\Psi = \int_{-\infty}^{\infty} \int_{-\infty}^{\infty} e^{-\frac{1}{2}\left(\frac{\tilde{c}_{x}}{\sigma_{x}}\right)^{2}-\frac{1}{2}\left(\frac{\tilde{c}_{z}-c_{z}}{\sigma_{z}}\right)^{2}}\sum_{m=0}^{3}\sum_{n=0}^{3} a_{mn}(\tilde{c}_{x})^{m}(\tilde{c}_{z}-c_{z})^{n} \frac{\tilde{c}_{z}}{\sqrt{\tilde{c}_{x}^{2}+\tilde{c}_{z}^{2}}} \mathrm{d}\tilde{c}_{x}\mathrm{d}\tilde{c}_{z}.
\end{equation}
Making the change to polar coordinates, $r=\sqrt{\tilde{c}_{x}^{2}+\tilde{c}_{z}^{2}}$, where $\tilde{c}_{z}=r\cos\theta$, $\tilde{c}_{x}=r\sin\theta$, and using the fact that $\sigma_{x}=\sigma_{z}$ we have
\begin{equation}
\Psi = e^{-\frac{c_{z}^{2}}{2\sigma_{z}^{2}}}\int_{0}^{\infty} \mathrm{d}r \, r^{1+m+n} e^{-\frac{r^2}{2\sigma_{z}^{2}}} \int_{0}^{2\pi} \mathrm{d}\theta \,e^{\frac{rc_{z}\cos\theta}{\sigma_{z}^{2}}} \sum_{m=0}^{3}\sum_{n=0}^{3}a_{mn}(\sin\theta)^{m}\left(\cos\theta - \frac{c_{z}}{r}\right)^{n}\cos\theta.
\end{equation}
Making the change of variables, $\alpha=\frac{rc_{z}}{\sigma_{z}^{2}}$, and noting that $\frac{c_{z}}{r}=\frac{c_{z}^{2}}{\sigma_{z}^{2}\alpha}=\frac{4s}{\alpha}$ we define the angular integration
\begin{equation}
\mathcal{I}_{mn}(\alpha,s) = \int_{0}^{2\pi}e^{\alpha \cos \theta} (\sin\theta)^{m} \left(\cos \theta - \frac{4s}{\alpha}\right)^{n} \cos \theta \,\mathrm{d}\theta.
\end{equation}
We compute the angular integrals associated with the non-zero $a_{mn}$, finding 
\begin{eqnarray}
\mathcal{I}_{00}(\alpha,s) &=& 2\pi I_{1}(\alpha) \\
\mathcal{I}_{10}(\alpha,s) &=& 0 \\
\mathcal{I}_{30}(\alpha,s) &=& 0 \\
\mathcal{I}_{01}(\alpha,s) &=& 2\pi \left( (1-4s)\frac{I_{1}(\alpha)}{\alpha} + I_{2}(\alpha) \right) \\
\mathcal{I}_{21}(\alpha,s) &=& 2\pi \left( \frac{I_{1}(\alpha)}{\alpha} -\left(3 + 4s\right)\frac{I_{2}(\alpha)}{\alpha^2} \right) \\
\mathcal{I}_{12}(\alpha,s) &=& 0 \\
\mathcal{I}_{03}(\alpha,s) &=& 2\pi \left( -4s\left(   \frac{3}{\alpha} + \frac{4s(4s-3)}{\alpha^3} \right)I_{1}(\alpha) + \left( 1 + \frac{12s(1+4s) +3}{\alpha^2} \right)I_{2}(\alpha)\right).
\end{eqnarray}
Here $I_{1}(\alpha)$ and $I_{2}(\alpha)$ are the modified Bessel functions of the first kind of order 1 and 2, respectively.  Defining the radial integration
\begin{equation}
\mathcal{R}_{mn}(s) = \int_{0}^{\infty}   \alpha^{1+m+n}e^{-\frac{\alpha^{2}}{8s}}\mathcal{I}_{mn}(\alpha,s) \mathrm{d}\alpha 
\end{equation}
we arrive at the result quoted in the main text
\begin{eqnarray}
\Psi &=& \sum_{m=0}^{3}\sum_{n=0}^{3}\Psi_{mn} \\
\Psi_{mn} &=& \left( \frac{\sigma_{z}^{2}}{c_{z}}\right)^{2+m+n}e^{-2s}a_{mn}\mathcal{R}_{mn}(s).
\end{eqnarray}
The results for the non-zero radial integrals can be written as, 
\begin{eqnarray}
\mathcal{R}_{00}(s) &=& 4\sqrt{2}(\pi s)^{\frac{3}{2}}e^{s}(I_{0}(s)+I_{1}(s))\\
\mathcal{R}_{01}(s) &=& 4\sqrt{2}(\pi s)^{\frac{3}{2}}e^{s}(I_{0}(s)-I_{1}(s)) \\
\mathcal{R}_{21}(s) &=& 24\sqrt{2}(\pi s)^{\frac{3}{2}}e^{s} I_{1}(s) \\
\mathcal{R}_{03}(s) &=& 8\sqrt{2} (\pi s)^{\frac{3}{2}}e^{s} (3+4s)(2sI_{0}(s)-(1+2s)I_{1}(s)).
\end{eqnarray}
Collecting these results a bit of algebra yields the contributions to the chemotactic index.  Defining the dimensionless ratio of concentrations, $\lambda = \frac{c_{z}}{c_{\infty}}$, we have
\begin{eqnarray}
\Psi_{00} &=& \left(\frac{\pi s}{2}\right)^{\frac{1}{2}}e^{-s} (I_{0}(s)+I_{1}(s))  \label{psiresult1} \\ 
\Psi_{01} &=& -\frac{1}{120} \frac{\lambda(10+54\lambda+45\lambda^{2}-27\lambda^{3})}{1-3\lambda^{2}} \left(\frac{2\pi}{s}\right)^{\frac{1}{2}}e^{-s}(I_{0}(s)-I_{1}(s))   \label{psiresult2} \\ 
\Psi_{21} &=& \frac{3}{80} \frac{\lambda(5+14\lambda + 3\lambda^{2}-18\lambda^{3})}{1-3\lambda^{2}}\left(\frac{\pi}{2s^{3}}\right)^{\frac{1}{2}}e^{-s}I_{1}(s)   \label{psiresult3} \\ 
\Psi_{03} &=& \frac{1}{160} \frac{\lambda(1+\lambda)(5+9\lambda)}{1-3\lambda^{2}} \left( \frac{2\pi}{s^{3}} \right)^{\frac{1}{2}}e^{-s} (3+4s)(2sI_{0}(s)-(1+2s)I_{1}(s)). \label{psiresult4}
\end{eqnarray}
\end{widetext}

\bibliography{cumulant}

\end{document}